\documentclass[aps,onecolumn,showpacs,showkeys,nofootinbib]{revtex4}
\usepackage{epsfig}
\usepackage{amsmath}
\usepackage{amsfonts}
\usepackage{amssymb}
\usepackage{graphics}
\usepackage{colordvi}

\begin{document}

\title{Conformal Transformations and  Weak Field Limit of Scalar-Tensor Gravity}

\author{A. Stabile$^{1}$\footnote{arturo.stabile@gmail.com},  An. Stabile$^{2,3}$\footnote{anstabile@gmail.com}, S. Capozziello$^{4,5}$\footnote{capozziello@na.infn.it}}
\affiliation{$^1$Dipartimento di Ingegneria, Universit\`{a} del Sannio, Palazzo Dell'Aquila Bosco Lucarelli, Corso Garibaldi, 107 - 82100, Benevento, Italy}
\affiliation{$^2$Dipartimento di Fisica "E.R. Caianiello", Universit\`{a} degli Studi di Salerno, via Ponte don Melillo, Stecca 9, I - 84084 Fisciano, Italy}
\affiliation{$^3$INFN Sezione di Napoli, Gruppo collegato di Salerno, via Ponte don Melillo, Stecca 9, I - 84084 Fisciano, Italy}
\affiliation{$^4$Dipartimento di Fisica, Universit\`{a} di Napoli "Federico II", Complesso Universitario di Monte Sant'Angelo, Edificio G, Via Cinthia, I-80126, Napoli, Italy}
\affiliation{$^5$INFN Sezione di Napoli, Complesso Universitario  di Monte Sant'Angelo, Edificio G, Via Cinthia, I-80126, Napoli, Italy}

\begin{abstract}
The weak field limit of scalar tensor theories of gravity is discussed in view of conformal transformations. Specifically, we consider how physical quantities, like gravitational potentials derived in the Newtonian approximation for the same scalar-tensor theory, behave in the Jordan and in the Einstein frame. The approach allows to discriminate features that are invariant  under  conformal transformations and gives contributions  in the debate of selecting the true physical frame. As a particular example, the case of $f(R)$ gravity is considered.

\end{abstract}
\pacs{conformal transformations; alternative theories of gravity; weak field limit}
\date{\today}
\maketitle

\section{Introduction}

The  current  cosmological observations   point out a spatially flat model with a bulk of  dark matter and  dark energy related to
a large negative pressure necessary to explain the observed accelerating
expansion of the Hubble fluid 
and the large scale structure \cite{Teg04,Spergel07,essence,Kowal08,Hic09,komatsu08,hin,kom,LJC09,jes,BasPli10}. Despite of the observational evidences,   the nature and
the  origin of dark sector remain  a non-solved puzzle  of  theoretical physics that give rise to 
 a plethora of alternative
cosmological scenarios. Most of them are based either on the existence of new fields (aimed to address the "dark" problem at fundamental level)
 or on extensions and  modifications of  General
Relativity. In this latter picture,   the accelerating behavior and the amount of dark matter can be seen as different
geometric effects  \cite{Ratra88,curvature,mauro,report,repsergei,
Oze87,Weinberg89,Lambdat,car,lim_1,lim_2,arb,ove,Bas09c,Wetterich:1994bg,
Caldwell98,Brax:1999gp,KAM,fein02,Caldwell,Bento03,chime04,Linder2004,LSS08,
Brookfield:2005td,Boehmer:2007qa,Starobinsky-2007,Ame10,cal,saw,ame,cap_fao,bam}.

From a genuine theoretical viewpoint, a straightforward way to
study dynamics  is to look for conformally related
models in order to disentangle further degrees of freedom. Such new degrees are not present in the standard view where only the Hilbert-Einstein action and perfect fluid matter are taken into account.   In particular, conformally equivalent theories can be used
 to select viable cosmological models \cite{allemandi}. This point has to be discussed in some detail.  In fact, by conformally transforming cosmological models can happen that some features as couplings and potentials can be directly related to the cosmological observables. The "selection" means that in a given conformal frame, some observational features are more evident. In \cite{allemandi}, examples in this sense are given. In particular, it is shown that several non-minimally coupled models, if conformally transformed, give rise to an effective cosmological constant and then can be directly matched with observations in the $\Lambda$CDM framework.

On the other hand,  further scalar fields (degrees of freedom) into the gravitational Lagrangian  
give rise to two separate classes of theories: minimally and non-minimally coupled theories. In  general, also higher-order theories of gravity can be reduced to the non-minimally coupled standard (see \cite{report} for details).

In the first case, the gravitational coupling is the  Newton constant. The  scalar fields 
are added to the Ricci scalar $R$  in the gravitational Lagrangian. In this case, we are dealing with the so-called {\it Einstein frame}.

In the second case,  the gravitational coupling is a a function of space and time and it is dynamically related to the scalar fields.  
The paradigm  is the  Brans-Dicke gravity, formerly deduced by Jordan which is closely related to what in later times got the name "Brans-Dicke gravity" \cite{brans,Jordan38,Jordan52}.  
It consists of  a scalar field $\phi$ non-minimally coupled to  $R$ and a kinetic term for 
the  scalar field into the gravitational action. As a result,  the coupling is non-minimal and 
 the gravitational interaction  changes with distance  and time according to the Mach principle. The straightforward generalization is  to take into account  theories  where also a  self-interacting potential or more scalar fields are  present. Furthermore,  gravitational theories non-linear in the Ricci scalar $R$ or containing other curvature invariants  can be reduced to  scalar-tensor ones.     
In general, when  we  take into account  non-minimal couplings or higher-order terms, we are dealing with  
the {\it Jordan frame}. 

The Einstein and Jordan frames are related by  geometrical maps that  are the {\it conformal
transformations} and the question is whether such frames are only mathematically equivalent or also physically equivalent. 
The problem of identifying the physical frame  has been longly debated  and  nowadays  
strongly emerges in order to address the problem of "dark sector"  either from a geometrical or a material viewpoint  \cite{basilakos}.

An important example is related to the geodesic motion.  In the Jordan frame, in vacuum, neutral massive test  particles fall along time-like geodesics. This is not true in the Einstein frame  where  they deviate from geodesic motion due to a
force coming from the conformal scalar field gradient. 
As a consequence, from conformal transformations point of view, the Equivalence Principle holds
only in the Jordan frame.   It is important to stress that such a  Principle is
the basic foundation of relativistic theories of gravity. 
 Then, a representation-independent formulation should physically discriminate between frames. 
 No final result holds  in this sense and the  violation of the Equivalence
Principle (in the Einstein frame) could be interpreted as the fact  that frames are not physically equivalent.
 On the other hand,  if the Equivalence 
Principle holds in a given frame and not in any frame means 
that it is not a covariant feature but only  a kinematical one. 
In other words, Equivalence Principle is not sufficient 
to discriminate between conformal  frames.

However,  the   vacuum interpretation has to be discussed.
It  have two different meanings: If the energy-momentum tensor is $T_{\mu\nu} =0$,
the scalar field belongs to
 the gravitational field sector  then it is a part of  geometry.  On the other hand,  if   $T_{\mu\nu} + T^\phi_{\mu\nu}=0$,  the sum of    matter fluid   and  scalar field  contributions is zero. In this case,   the scalar field can be considered 
 as a  matter field. In the Jordan frame,  
 both  interpretations have the same meaning as soon as  the scalar
 field gradient is zero. In other words, the contracted Bianchi identities must hold.  However, the meaning of  vacuum is different and then also the motion of  test particles  moving along  geodesics is different. This fact has to be carefully considered if one wants to discriminate between Einstein and Jordan frames.

Furthermore, there are results  where exact cosmological solutions accelerate in
one frame but not in the other. This fact could mean that, for an
astronomer attempting to fit observations,  the
two frames  are not  physically equivalent
\cite{nodicap2, CapozzielloPrado}.
In these  situations, one must state precisely what the
physical equivalence is and the concept is not obvious at all. In a  naive formulation, such an equivalence could be related to the fact that it should be possible to select a set of physically invariant quantities that can be conformally transformed.

As we said,  conformal transformations allow to disentangle the further gravitational degrees of freedom coming from general actions \cite{report, MagnanoSokolowski94}. The idea is to perform a conformal rescaling of the space-time metric $g_{\mu\nu}\rightarrow \tilde{g}_{\mu\nu}$ and a redefinition  of  the scalar field $\phi$ as  $\phi\,\rightarrow\,\tilde{\phi}$.  New dynamical variables $\left\{\tilde{g}_{\mu\nu}, \tilde{\phi} \right\}$ are thus obtained. The scalar field redefinition allows, for example,  to cast the kinetic energy density of this field in a canonical form. The new set of variables $\left\{\tilde{g}_{\mu\nu}, \tilde{\phi} \right\}$ defines  the \emph{Einstein conformal frame}, while $\left\{ g_{\mu\nu}, \phi \right\}$ constitutes the \emph{Jordan frame}. When a scalar degree of freedom $\phi$ is present in  the theory, as in scalar tensor or $f(R)$ gravity, it generates the transformation to  the Einstein frame  in the sense that the rescaling is completely determined by a function of $\phi$. In principle, infinite conformal frames could be introduced, giving rise to  many representations of the theory. 

Let the pair  $\{\mathcal{M}, g_{\mu\nu}\}$ be  a space-time, with ${\cal M}$ a smooth manifold of  dimension $n \geq 2$ and $g_{\mu\nu} $ a (pseudo)-Riemannian metric on $\mathcal{M}$. The point-dependent rescaling of the metric tensor

\begin{eqnarray}\label{transconf}
g_{\mu\nu}\, \longrightarrow\,\tilde{g}_{\mu\nu}\,=\,\Omega^2\,g_{\mu\nu}
\end{eqnarray}
where $\Omega\,=\,\Omega(x)$ is a nowhere vanishing, regular function, called a \emph{Weyl} or \emph{conformal} 
transformation. Obviously the transformation rule for the controvariant metric tensor is
$\tilde{g}^{\mu\nu}\,=\,\Omega^{-2}g^{\mu\nu}$.

Due to this metric rescaling, the lengths of space-like and time-like intervals and the norms of space-like and time-like vectors  change, while  null vectors and null intervals of the metric $g_{\mu\nu}$ remain null in the rescaled metric $\tilde{g}_{\mu\nu}$ (in this sense, they are conformally invariant quantities).  The light cones are left unchanged by the transformation (\ref{transconf}) and the space-times  $\{\mathcal{M}, g_{\mu\nu}\}$ and  $\{\mathcal{M}, \tilde{g}_{\mu\nu}\} $ exhibit the same causal structure; the converse is also true \cite{Wald84}. A vector that is time-like, space-like, or null with respect to the metric $g_{\mu\nu}$ has the same character with respect to $\tilde{g}_{\mu\nu}$, and \emph{vice-versa}.

In wide sense, conformal invariance corresponds to the absence of  characteristic lengths  and masses. In general, the effective potential of scalar field $V(\phi)$  coming from conformal transformations contains dimensional parameters  (such as  masses, that are  further "characteristic gravitational lengths"). This means that the further degrees of freedom coming from extended or alternative gravities give rise to features that could play a fundamental role in the dynamics of astrophysical structures, from the "infrared"  side, and in quantum gravity, from the "ultraviolet"  side. 

However, an important remark is necessary here. Typically,  the absence of characteristic lengths and masses
 is called {\it scale-invariance}, and even for scale-invariance, several
 different interpretations exist.  This means that conformal invariance and   scale-invariance must be precisely distinguished.
 As discussed, for example,  in \cite{singleton}, a class of isotropic cosmologies in fourth-order gravity with Lagrangians of the form ${\cal L} = F(R)+K({\cal G})$, where $R$ and ${\cal G}$ are the Ricci and Gauss-Bonnet scalars respectively, can be made  {\it scale-invariant}. It is important to stress that such theories can be also conformally transformed.  This is a typical case where the two invariances can be clearly distinguished.  In general, scale invariance means that  physical systems  do not change if scales of length, energy, or other variables, are multiplied by some factor. Technically  this transformation is a  {\it dilation}.  Such a feature can be part of a larger conformal symmetry where angles are preserved.

In this paper, we want to address the problem of how conformally transformed models behave in the weak field limit approximation. This issue could be  extremely relevant in order to select conformally invariant physical quantities. 

This point deserves some discussion.
In general,  a gauge theory is a type of field theory where the Lagrangian is invariant under a continuous group of local transformations. In particular, gravitation is a  field theory on a principal frame bundle whose gauge symmetries are covariant transformations \cite{foop}. In this case, 
the term "gauge" refers to redundant degrees of freedom in the Lagrangian. The transformations between possible gauges, called gauge transformations, form a Lie group which is  the symmetry group or the gauge group of the theory.

On the other hand, the conformal group is the group of transformations from a space to itself that preserve all angles within the space. More formally, it is the group of transformations that preserve the conformal geometry of the space. These definitions immediately point out that the gauge and conformal  groups do not coincide and then breaking gauge invariance could be not related to conformal invariance.

Furthermore,  gauge invariance is  broken in the weak field limit approximation and redundant degrees of freedom can be gauged away by this procedure. Comparing two conformally related models in the weak field limit  could be a procedure to select physically invariant quantities once the behavior of gauges  in the two frames is  determined and their conformal transformations derived.

With these considerations in mind,  we will take into account the weak field limit of scalar-tensor gravity in the Jordan frame (Sec. II) and compare it with the analogous in the Einstein frame (Sec. III). The particular case of $f(R)$ gravity will be considered in Sec. IV. Discussion and conclusions are drawn in Sec.V.

\section{Scalar tensor gravity in the Jordan frame}

The action of a  scalar tensor theory of gravity in 4 dimensions is
\begin{eqnarray}\label{ST_action_1}
\mathcal{A}^{JF}\,=\,\int d^{4}x\sqrt{-g}\biggl[\phi\,R+V(\phi)+\omega(\phi)\,\phi_{;\alpha}\,\phi^{;\alpha}+\mathcal{X}\mathcal{L}_m\biggr]
\end{eqnarray}
where $R$ is the Ricci scalar and $\mathcal{X}\,=\,8\pi G$ where we assumed
$c\,=\,1$. The convention for the Ricci tensor is
$R_{\mu\nu}={R^\sigma}_{\mu\sigma\nu}$, while for the Riemann
tensor is ${R^\alpha}_{\beta\mu\nu}=\Gamma^\alpha_{\beta\nu,\mu}+...$. The
affinities are the standard Christoffel symbols of the metric:
$\Gamma^\mu_{\alpha\beta}=\frac{1}{2}g^{\mu\sigma}(g_{\alpha\sigma,\beta}+g_{\beta\sigma,\alpha}
-g_{\alpha\beta,\sigma})$. The adopted signature is $(+---)$ while the coordinates $x^\mu\,=\,(x^0,x^1,x^2,x^3)\,=\,(t,\textbf{x})$ are the  \emph{isotropic coordinates}. The Greek index runs from $0$ to $3$; the Latin index runs from $1$ to $3$.
Let us  note that the action
\begin{equation}
\mathcal{A}^{JF}\,=\,\int d^{4}x\sqrt{-g}\biggl[F(\phi)R+V(\phi)+\omega(\phi)\,\phi_{;\alpha}\,\phi^{;\alpha}+\mathcal{X}\mathcal{L}_m\biggr]
\end{equation}
is apparently more general than (\ref{ST_action_1}). In fact by substituiting $F(\phi)\,\rightarrow\,\phi$, we obtain only a new definition of functions $\omega(\phi)$ and $V(\phi)$ so the two formulations are essentially equivalent.

The term $\mathcal{L}_m$ is the minimally coupled ordinary matter contribution  considered as a \emph{perfect fluid}; $\omega(\phi)$ is a function of the scalar field and $V(\phi)$ is its potential which specifies  the dynamics. Actually if  $\omega(\phi) =\pm 1,0$ the nature and the dynamics of the scalar field is fixed. It can be a canonical scalar field, a phantom field or a field without dynamics (see e.g. \cite{odi2005,singularity} for details). In the metric approach, the field equations are obtained by varying the action (\ref{ST_action_1}) with respect to $g_{\mu\nu}$ and $\phi$. The field equations are 

\begin{eqnarray}
\label{fieldequation_ST}
&&\phi\,R_{\mu\nu}-\frac{\phi\,R+V(\phi)+\omega(\phi)\,\phi_{;\alpha}\,\phi^{;\alpha}}{2}\,g_{\mu\nu}+\omega(\phi)\,\phi_{;\mu}\,\phi_{;\nu}-\phi_{;\mu\nu}+g_{\mu\nu}\Box\,
\phi\,=\,\mathcal{X}\,T_{\mu\nu}\nonumber\\\\
&&2\,\omega(\phi)\,\Box\,\phi+\omega_{\phi}(\phi)\,\phi_{;\alpha}\phi^{;\alpha}-R-V_{\phi}(\phi)\,=\,0\nonumber
\end{eqnarray}
and the trace equation  is 
\begin{equation}
\phi\,R+2V(\phi)+\omega(\phi)\,\phi_{;\alpha}\phi^{;\alpha}-3\,\Box\,\phi\,=\,-\mathcal{X}\,T
\end{equation}
Here we introduced, respectively, the energy-momentum tensor of matter and the d'Alembert operator

\begin{eqnarray}\label{en_ten}
T_{\mu\nu}\,=\,-\frac{1}{\sqrt{-g}}\frac{\delta(\sqrt{-g}\,\mathcal{L}_m)}{\delta
g^{\mu\nu}}\,,\,\,\,\,\,\,\,\,\,\,\Box\,(\cdot)\,=\,\frac{\partial_\sigma(\sqrt{-g}\,g^{\sigma\tau}\partial_\tau(\cdot))}{\sqrt{-g}}
\end{eqnarray}
$T\,=\,T^{\sigma}_{\,\,\,\,\,\sigma}$ is the trace of energy-momentum tensor and $V_\phi\,=\,\frac{dV}{d\phi}$, $\omega_\phi(\phi)\,=\,\frac{d\omega(\phi)}{d\phi}$. 
If we assume that the Lagrangian density $\mathcal{L}_m$ of  matter depends only on the metric components $g_{\mu\nu}$ and not on its derivatives, we obtain $T_{\mu\nu}\,=\,1/2\,\mathcal{L}_m\,g_{\mu\nu}-\delta\mathcal{L}_m/\delta g^{\mu\nu}$. Let us consider a source with mass $M$. The energy-momentum tensor is 
\begin{eqnarray}\label{emtensor}
T_{\mu\nu}\,=\,\rho\,u_\mu
u_\nu\,,\,\,\,\,\,\,\,\,\,\,T\,=\,\rho
\end{eqnarray}
where $\rho$ is the mass density, $u_\mu$ satisfies the condition $g^{00}{u_0}^2\,=\,1$, and $u_i\,=\,0$.  Here, we are not interested  to the internal structure. It is useful to get the expression of $\mathcal{L}_m$. In fact from the definition (\ref{en_ten}),  we have

\begin{eqnarray}
\delta\int d^4x\sqrt{-g}\,\mathcal{L}_m\,=\,-\int d^4x\sqrt{-g}\,T_{\mu\nu}\,\delta g^{\mu\nu}\,=\,-\int d^4x\sqrt{-g}\,\rho\,u_\mu u_\nu\,\delta g^{\mu\nu}
\end{eqnarray}
From the mathematical properties of metric tensor we have 
\begin{equation}
\delta(\sqrt{-g}\,\rho)\,=\,1/2\,\sqrt{-g}\,\rho\,u^\mu u^\nu\,\delta g_{\mu\nu}\,=\,-1/2\,\sqrt{-g}\,\rho\,u_\mu u_\nu\,\delta g^{\mu\nu}
\end{equation}
 then we find
\begin{eqnarray}
\mathcal{L}_m\,=\,2\,\rho
\end{eqnarray}
The variation of density is given by 
\begin{equation}
\delta\rho\,=\,\frac{\rho}{2}(g_{\mu\nu}-u_\mu u_\nu)\,\delta g^{\mu\nu}
\end{equation}
 order to deal with standard self-gravitating systems, any theory of gravity has to be developed in its Newtonian or post-Newtonian limit depending on the order of approximation  in terms of squared velocity $v^2$  \cite{cqg,stabile}. The  Newtonian limit starts from  developing  the
metric tensor (and other additional quantities in the theory) with respect to the dimensionless velocity\footnote{The velocity $v$ is here expressed in light speed units.} $v$ of the moving massive bodies embedded in the gravitational potential. The perturbative development takes  only first term of $(0,0)$- and $(i,j)$-component of metric tensor $g_{\mu\nu}$ (for details, see \cite{stabile,stabile2}). The  metric assumes the  form

\begin{eqnarray}\label{me_JF}
{ds}^2\,=\,(1+2\Phi)\,dt^2-(1-2\Psi)\,\delta_{ij}dx^idx^j
\end{eqnarray}
where the gravitational potentials $\Phi,\, \Psi\,<\,1$ are proportional to $v^2$. The Ricci scalar is approximated as $R\,=\,R^{(1)}\,+\,R^{(2)}\,+\,\dots$ where $R^{(1)}$ is proportional to $\Phi$, and $\Psi$, while $R^{(2)}$ is proportional to $\Phi^2$, $\Psi^2$ and $\Phi\Psi$. In this context,  also the scalar field $\phi$ is approximated as the Ricci scalar. In particular we get $\phi\,=\,\phi^{(0)}\,+\,\phi^{(1)}\,+\dots$ while the functions $V(\phi)$ and $\omega(\phi)$ can be substituted by their corresponding Taylor series.

From the lowest order of field Eqs. (\ref{fieldequation_ST}) we have

\begin{eqnarray}\label{PPN-field-equation-general-theory-ST-O0}
V(\phi^{(0)})\,=\,0\,,\,\,\,\,\,\,\,\,\,\,V_{\phi}(\phi^{(0)})\,=\,0
\end{eqnarray}
and also in the scalar tensor gravity a missing cosmological component in the action (1) implies that the space-time is asymptotically Minkowskian; moreover the ground value of scalar field $\phi$ must be a stationary point of potential. In the Newtonian limit,  we have

\begin{eqnarray}
\label{NL-fieldequation_ST}
&&\triangle\biggl[\Phi-\frac{\phi^{(1)}}{\phi^{(0)}}\biggr]-\frac{R^{(1)}}{2}\,=\,\frac{\mathcal{X}\,\rho}{\phi^{(0)}}\nonumber\\\nonumber\\
&&\biggl\{\triangle\biggl[\Psi+\frac{\phi^{(1)}}{\phi^{(0)}}\biggr]+\frac{R^{(1)}}{2}\biggr\}\delta_{ij}+\biggr\{\Psi-\Phi-\frac{\phi^{(1)}}{\phi^{(0)}}\biggr\}_{,ij}\,=\,0\nonumber\\\\
&&\triangle\phi^{(1)}+\frac{V_{\phi\phi}(\phi^{(0)})}{2\,\omega(\phi^{(0)})}\,\phi^{(1)}+\frac{R^{(1)}}{2\,\omega(\phi^{(0)})}\,=\,0\nonumber\\\nonumber\\
&&R^{(1)}+3\,\frac{\triangle\phi^{(1)}}{\phi^{(0)}}\,=\,-\frac{\mathcal{X}\,\rho}{\phi^{(0)}}\nonumber
\end{eqnarray}
where $\triangle$ is the Laplacian in the flat space. These equations are not simply the merging of field equations of GR and a further massive scalar field, but  come out to the fact that the scalar tensor gravity generates a coupled system of equations with respect to Ricci scalar $R$ and scalar field $\phi$. The gravitational potentials $\Phi$, $\Psi$ and the Ricci scalar $R^{(1)}$ are given by
\begin{eqnarray}\label{new_sol}
&&\Phi(\mathbf{x})\,=\,-\frac{\mathcal{X}}{4\pi\,\phi^{(0)}}\int
d^3\textbf{x}'\frac{\rho(\textbf{x}')}{|\textbf{x}-
\textbf{x}'|}-\frac{1}{8\pi}\int
d^3\textbf{x}'\frac{R^{(1)}(\textbf{x}')}{|\textbf{x}-
\textbf{x}'|}+\frac{\phi^{(1)}(\textbf{x})}{\phi^{(0)}}\nonumber\\\nonumber\\
&&\Psi(\mathbf{x})\,=\,\Phi(\textbf{x})+\frac{\phi^{(1)}(\textbf{x})}{\phi^{(0)}}\\\nonumber\\
&&R^{(1)}(\textbf{x})\,=\,-\frac{\mathcal{X}\,\rho(\textbf{x})}{\phi^{(0)}}-3\,\frac{\triangle\phi^{(1)}(\textbf{x})}{\phi^{(0)}}\nonumber
\end{eqnarray}
and supposing that $2\,\omega(\phi^{(0)})\,\phi^{(0)}-3\,\neq\,0$ we find for the scalar field $\phi^{(1)}$ the Yukawa-like field equation

\begin{eqnarray}
\label{fieldequation_SF}
\biggl[\triangle-{m_\phi}^2\biggr]\phi^{(1)}\,=\,\frac{\mathcal{X}\,\rho}{2\,\omega(\phi^{(0)})\phi^{(0)}-3}
\end{eqnarray}
where we introduced the mass definition

\begin{eqnarray}\label{mass_defin}
{m_\phi}^2\,\doteq\,-\frac{\phi^{(0)}\,V_{\phi\phi}(\phi^{(0)})}{2\,\omega(\phi^{(0)})\,\phi^{(0)}-3}\,.
\end{eqnarray}
It is important to stress that the potential $\Psi$ can be found also as  
\begin{equation}
\Psi(\textbf{x})\,=\,\frac{1}{8\pi}\int
d^3\textbf{x}'\frac{R^{(1)}(\textbf{x}')}{|\textbf{x}-
\textbf{x}'|}-\frac{\phi^{(1)}(\textbf{x})}{\phi^{(0)}}
\end{equation}
see for example  \cite{sta_cap}.

By using the Fourier transformation, the solution of Eq. (\ref{fieldequation_SF}) has the following form

\begin{eqnarray}
\label{fieldequation_SF_sol}
\phi^{(1)}(\textbf{x})\,=\,-\frac{\mathcal{X}}{2\,\omega(\phi^{(0)})\phi^{(0)}-3}\int\frac{d^3\textbf{k}}{(2\pi)^{3/2}}\frac{\tilde{\rho}(\textbf{k})\,e^{i\textbf{k}\cdot\textbf{x}}}{\textbf{k}^2+{m_\phi}^2}
\end{eqnarray}
The expressions (\ref{new_sol}) and (\ref{fieldequation_SF_sol}) represent the most general solution of any scalar-tensor gravity in the Newtonian limit. Since the superposition  principle is yet valid (the field Eqs. (\ref{NL-fieldequation_ST}) are linear), it  is sufficient to consider the solutions generated by a point-like source with mass $M$.  Then if we consider $\rho\,=\,M\,\delta(\textbf{x})$ the solutions are \cite{stabile,stabile2,sta_cap} 

\begin{eqnarray}
\label{NL-solution_ST}
&&\phi^{(1)}(\textbf{x})\,=\,-\frac{1}{2\,\omega(\phi^{(0)})\,\phi^{(0)}-3}\frac{r_g}{|\textbf{x}|}\,e^{-m_\phi |\textbf{x}|}\nonumber\\\nonumber\\
&&R^{(1)}(\textbf{x})\,=\,-\frac{4\pi\,r_g}{\phi^{(0)}}\,\delta(\textbf{x})+\frac{3\,{m_\phi}^2}{[2\,\omega(\phi^{(0)})\,\phi^{(0)}-3]\,\phi^{(0)}}\frac{r_g}{|\textbf{x}|}\,e^{-m_\phi |\textbf{x}|}\nonumber\\\\
&&\Phi(\textbf{x})\,=\,-\frac{GM}{\phi^{(0)}|\textbf{x}|}\biggl\{1-\frac{e^{-m_\phi |\textbf{x}|}}{2\,\omega(\phi^{(0)})\,\phi^{(0)}-3}\biggr\}\nonumber\\\nonumber\\
&&\Psi(\textbf{x})\,=\,-\frac{GM}{\phi^{(0)}|\textbf{x}|}\biggl\{1+\frac{e^{-m_\phi |\textbf{x}|}}{2\,\omega(\phi^{(0)})\,\phi^{(0)}-3}\biggr\}\nonumber
\end{eqnarray}
where $r_g\,=\,2GM$ is the Schwarzschild radius. In the case $V(\phi)\,=\,0$, the scalar field is massless and $\omega(\phi)\,=\,-\omega_0/\phi$, we obtain

\begin{eqnarray}\label{sol_point_BD}
\Phi(\textbf{x})\,=\,\Phi_{BD}(\mathbf{x})\,&=&\,-\frac{GM}{\phi^{(0)}|\textbf{x}|}\left[\frac{2(2+\omega_0)}{2\,\omega_0+3}\right]
\,=\,-\frac{G^*M}{|\textbf{x}|}
\nonumber\\\\
\Psi(\textbf{x})\,=\,\Psi_{BD}(\mathbf{x})\,&=&\,-\frac{G^*M}{|\textbf{x}|}\left(\frac{1+\omega_0}{2+\omega_0}\right)
\nonumber
\end{eqnarray}
the well-known  Brans-Dicke solutions \cite{brans} with Eddington's parameter $\gamma\,=\,\frac{1+\omega_0}{2+\omega_0}$ \cite{will} where   the gravitational constat is defined as $G\,\rightarrow\,G^*\,=\,\frac{G}{\phi^{(0)}}\frac{2(2+\omega_0)}{2\,\omega_0+3}$.

\section{Scalar tensor gravity in the Einstein frame}

Let us now introduce the conformal transformation (\ref{transconf}) to show that scalar-tensor theories are, in general,   conformally equivalent to the Einstein theory plus minimally coupled scalar fields. However if standard matter is present, the conformal transformation generates the non-minimal coupling between the matter component and the scalar field.

By applying the transformation (\ref{transconf}), the action in (\ref{ST_action_1}) can be
reformulated as follows

\begin{eqnarray}\label{TS-EF-action}
\mathcal{A}^{EF}\,=\,\int d^4x \sqrt{-\tilde{g}}
\biggl[\Xi\,\tilde{R}+W(\tilde{\phi})
+\tilde{\omega}(\tilde{\phi})\tilde{\phi}_{;\alpha}\tilde{\phi}^{;\alpha}+\mathcal{X}
\tilde{\mathcal{L}}_{m}\biggr]
\end{eqnarray}
in which $\tilde{R}$ is the Ricci scalar relative to the metric
$\tilde{g}_{\mu\nu}$ and $\Xi$ is a generic constant. 
The two actions (\ref{ST_action_1}) and (\ref{TS-EF-action}) are mathematically equivalent. In fact the conformal transformation is given by imposing the condition 
\begin{equation}
\sqrt{-g}\biggl[\phi\,R+V(\phi)+\omega(\phi)\,\phi_{;\alpha}\,\phi^{;\alpha}+\mathcal{X}\mathcal{L}_m\biggr]\,=\,\sqrt{-\tilde{g}}
\biggl[\Xi\,\tilde{R}+W(\tilde{\phi})+\tilde{\omega}(\tilde{\phi})\tilde{\phi}_{;\alpha}\tilde{\phi}^{;\alpha}+\mathcal{X}\tilde{\mathcal{L}}_{m}\biggr]
\end{equation}
The relations
between the quantities in the two frames are

\begin{eqnarray}\label{transconfTS}
&&\tilde{\omega}(\tilde{\phi})\,{d\tilde{\phi}}^2\,=\,\frac{\Xi}{2}\,[2\,\phi\,\omega(\phi)-3]\biggl(\frac{d\phi}{\phi}\biggr)^2\nonumber\\\nonumber\\
&&W(\tilde{\phi})=\frac{\Xi^2}{\phi(\tilde{\phi})^2}\,V(\phi(\tilde{\phi}))\nonumber\\\\
&&\tilde{\mathcal{L}}_m\,=\,\frac{\Xi^2}{\phi(\tilde{\phi})^2}\,\mathcal{L}_m\biggl(\frac{\Xi\,\tilde{g}_{\rho\sigma}}{\phi(
\tilde{\phi})}\biggr)\nonumber\\\nonumber\\
&&\phi\,\Omega^{-2}\,=\,\Xi\nonumber
\end{eqnarray}
The field equations for the new fields $\tilde{g}_{\mu\nu}$ and
$\tilde{\phi}$ are

\begin{eqnarray}\label{fieldequation_ST_EF}
&&\Xi\,\tilde{R}_{\mu\nu}-\frac{\Xi\,\tilde{R}+W(\tilde{\phi})+\tilde{\omega}(\tilde{\phi})\,\tilde{\phi}_{;\alpha}\tilde{\phi}^{;\alpha}}{2}\,\tilde{g}_{\mu\nu}+\tilde{\omega}(\tilde{\phi})\,\tilde{\phi}_{;\mu}\tilde{\phi}_{;\nu}\,=\,\mathcal{X}\,\tilde{T}_{\mu\nu}\nonumber
\\\nonumber\\
&&2\,\tilde{\omega}(\tilde{\phi})\,\tilde{\Box}\tilde{\phi}+\tilde{\omega}_{\tilde{\phi}}(\tilde{\phi})\,\tilde{\phi}_{;\alpha}\tilde{\phi}^{;\alpha}-W_{\tilde{\phi}}(\tilde{\phi})-\mathcal{X}\frac{\delta\,\tilde{\mathcal{L}}_m}{\delta\,\tilde{\phi}}\,=\,0
\\\nonumber\\
&&\Xi\,\tilde{R}+2W(\tilde{\phi})+\tilde{\omega}(\tilde{\phi})\,\tilde{\phi}_
{;\alpha}\tilde{\phi}^{;\alpha}\,=\,-\mathcal{X}\,\tilde{T}\nonumber
\end{eqnarray}
where $\tilde{T}_{\mu\nu}$ and $\tilde{\Box}$ are the re-definition of the quantities (\ref{en_ten}) with respect to the metric $\tilde{g}_{\mu\nu}$. The field Eqs. (\ref{fieldequation_ST_EF}) can be obtained from (\ref{fieldequation_ST}) by substituing all geometrical and physical quantities in terms of conformally transformed ones. In particular we have

\begin{eqnarray}\label{transf_conf_II}
&&R_{\mu\nu}\,=\,\tilde{R}_{\mu\nu}+2\,{\ln \Omega}_{\tilde{;\mu\nu}}+2\,{\ln \Omega}_{;\mu}\,{\ln \Omega}_{;\nu}+[\tilde{\Box}\ln\Omega-2\,\tilde{{\ln \Omega}^{;\sigma}{\ln \Omega}_{;\sigma}}]\,\tilde{g}_{\mu\nu}\nonumber\\\nonumber\\
&&R\,=\,\Omega^2\biggl[\tilde{R}+6\,\tilde{\Box}\ln\Omega-3\,\tilde{{\ln \Omega}^{;\sigma}{\ln \Omega}_{;\sigma}}\biggr]\nonumber
\\\\
&&\phi_{;\mu\nu}\,=\,\phi_{\tilde{;\mu\nu}}+2\,\phi_{;\mu}\phi_{;\nu}-\tilde{{\ln \Omega}^{;\sigma}\phi_{;\sigma}}\,\tilde{g}_{\mu\nu}\nonumber\\\nonumber\\
&&\Box(\cdot)\,=\,\Omega^2\,\tilde{\Box}(\cdot)-2\,\tilde{{\ln \Omega}^{;\sigma}\partial_{\sigma}}(\cdot)\nonumber
\end{eqnarray}

The integration of field Eqs. (\ref{fieldequation_ST_EF}) is only formal because we do not know the analytical expression of the coupling function between the matter and the scalar field $\tilde{\phi}$ (see the third line of (\ref{transconfTS})). We can make some assumptions on the  parameter $\Xi$ and the function $\tilde{\omega}(\tilde{\phi})$ in the minimally coupled Lagrangian (\ref{TS-EF-action}) and on the function $\omega(\phi)$ in the  nonminimally coupled Lagrangian (\ref{ST_action_1}). If we choose $\tilde{\omega}(\tilde{\phi})\,=\,-1/2$, $\Xi\,=\,1$ and $\omega(\phi)\,=\,-\omega_0/\phi$, the transformation between  the  scalar fields $\phi$ and  $\tilde{\phi}$ is given by the first line in (\ref{transconfTS}), that is 

\begin{eqnarray}\label{trans_rule}
\tilde{\phi}(\phi)\,=\,\tilde{\phi}_0+\sqrt{2\omega_0+3}\,\ln\phi\,\,\,\,\,\,\,\,\,\,\,\,\,\,\,\phi(\tilde{\phi})\,=\exp{\left(\frac{\tilde{\phi}-\tilde{\phi}_0}{\sqrt{2\omega_0+3}}\right)}
\end{eqnarray} 
where obviously $\omega_0\,>\,-3/2$ and $\tilde{\phi}_0$ is an integration constant\footnote{Without losing  generality, we can set $\tilde{\phi}_0\,=\,0$.}. The potential $W$ and the matter Lagrangian $\tilde{\mathcal{L}}_m$ are

\begin{eqnarray}\label{transconfTS_1}
W(\tilde{\phi})\,=\exp{\left(-\frac{2\tilde{\phi}}{\sqrt{2\omega_0+3}}\right)}\,V\biggl(e^{\frac{\tilde{\phi}}{\sqrt{2\omega_0+3}}}\biggr)\,\,\,\,\,\,\,\,\,\,\,\,\,\,\
\tilde{\mathcal{L}}_m\,=\,2\,\rho\exp{\left(-\frac{2\tilde{\phi}}{\sqrt{2\omega_0+3}}\right)}
\end{eqnarray}

In both frames, the scalar fields are expressed as perturbative contributions on the cosmological background ($\phi^{(0)}$, $\tilde{\phi}^{(0)}$) with respect to  the dimensionless quantity $v^2$. Then also for the scalar field $\tilde{\phi}$, we can consider the develop $\tilde{\phi}\,=\,\tilde{\phi}^{(0)}+\tilde{\phi}^{(1)}+\dots$. Such a develop can be applied to the transformation rule (\ref{trans_rule}) and we obtain  

\begin{eqnarray}\label{trans_rule_pert}
&&\tilde{\phi}(\phi)\,=\,\sqrt{2\omega_0+3}\,\ln\phi\,=\,\sqrt{2\omega_0+3}\,\ln\phi^{(0)}+\frac{\sqrt{2\omega_0+3}}{\phi^{(0)}}\,\phi^{(1)}+\dots\,\doteq\,\tilde{\phi}^{(0)}+\tilde{\phi}^{(1)}\,+\dots\nonumber\\\\\nonumber
&&\phi(\tilde{\phi})\,=\,e^{\frac{\tilde{\phi}}{\sqrt{2\omega_0+3}}}\,=\,e^{\frac{\tilde{\phi}^{(0)}}{\sqrt{2\omega_0+3}}}\,+\,\frac{e^{\frac{\tilde{\phi}^{(0)}}{\sqrt{2\omega_0+3}}}}{\sqrt{2\omega_0+3}}\,\tilde{\phi}^{(1)}\,+\dots\,\doteq\,\phi^{(0)}+\phi^{(1)}\,+\,\dots
\end{eqnarray} 
Since we are interested in the Newtonian limit of field Eqs. (\ref{fieldequation_ST_EF}), we can assume, for the conformally transformed metric $\tilde{g}_{\mu\nu}$, an expression as (\ref{me_JF}) but with  some differences. In fact from the conformal transformation ($\ref{transconf}$) and from the last line of (\ref{transconfTS}), we have

\begin{eqnarray}\label{transconf_NL}\tilde{g}_{\mu\nu}\,=\,\phi\,g_{\mu\nu}\,=\,\phi^{(0)}\eta_{\mu\nu}+[\phi^{(0)}g^{(1)}_{\mu\nu}+\phi^{(1)}\eta_{\mu\nu}]+\dots\,=\,\tilde{\eta}_{\mu\nu}+\tilde{g}^{(1)}_{\mu\nu}+\dots\end{eqnarray}
then the conformally transformed metric becomes
 
\begin{eqnarray}\label{me_EF}
{ds}^2\,=\,(\phi^{(0)}+2\tilde{\Phi})\,dt^2-(\phi^{(0)}-2\tilde{\Psi})\,\delta_{ij}dx^idx^j
\end{eqnarray}
and the relation between the gravitational potentials in the two frames is

\begin{eqnarray}\label{diff_pot}
\tilde{\Phi}-\phi^{(0)}\,\Phi\,=\,\frac{\phi^{(1)}}{2}\,,\,\,\,\,\,\,\,\,\,\,\,\,\,\,\,\tilde{\Psi}-\phi^{(0)}\,\Psi\,=\,-\frac{\phi^{(1)}}{2}
\end{eqnarray}
 Then the field Eqs. (\ref{fieldequation_ST_EF}) become\footnote{With the assumptions of the metric (\ref{me_EF}) the Ricci tensor $\tilde{R}_{\mu\nu}$ in the Newtonian limit has the form $\frac{\triangle\tilde{\Phi}}{\phi^{(0)}}$ (a similar behaviour for $\tilde{R}^{(1)}_{ij}$), where the Ricci scalar is scaled by the factor ${\phi^{(0)}}^2$. The same scaling occurs for the Laplacian: $\triangle\,\rightarrow\,\frac{\triangle}{\phi^{(0)}}$.}

\begin{eqnarray}\label{NL_fieldequation_ST_EF}
&&\frac{\triangle\tilde{\Phi}}{\phi^{(0)}}-\frac{\tilde{R}^{(1)}}{2}\,\phi^{(0)}\,=\,\mathcal{X}\,\tilde{T}^{(1)}_{00}\nonumber
\\\nonumber\\
&&\biggl\{\frac{\triangle\tilde{\Psi}}{\phi^{(0)}}+\frac{\tilde{R}^{(1)}}{2}\phi^{(0)}\biggr\}\,\delta_{ij}+\frac{(\tilde{\Psi}-\tilde{\Phi})_{,ij}}{\phi^{(0)}}\,=\,0\nonumber
\\\\
&&\frac{\triangle\tilde{\phi}^{(1)}}{\phi^{(0)}}-W_{\tilde{\phi}\tilde{\phi}}(\tilde{\phi}^{(0)})\,\tilde{\phi}^{(1)}-\mathcal{X}\biggl[\frac{\delta\,\tilde{\mathcal{L}}_m}{\delta\,\tilde{\phi}}\biggr]^{(1)}\,=\,0
\nonumber\\\nonumber\\
&&\tilde{R}^{(1)}\,=\,-\mathcal{X}\,\tilde{T}^{(1)}\nonumber
\end{eqnarray}
where also in this case we have $W(\tilde{\phi}^{(0)})\,=\,0$ and $W_{\tilde{\phi}}(\tilde{\phi}^{(0)})\,=\,0$.  However these conditions are an obvious consequence of the conformal transformation of conditions $V(\phi^{(0)})\,=\,0$ and $V_\phi(\phi^{(0)})\,=\,0$. In fact we can figure out  that $V(\phi)\,\propto\,(\phi-\phi^{(0)})^2$ and then $W(\tilde{\phi})\,\propto\,\biggl(e^{\frac{\tilde{\phi}}{\sqrt{2\omega_0+3}}}-\phi^{(0)}\biggr)^2$ which, by using  relations (\ref{trans_rule_pert}), satisfies the above conditions. Finally, we note that $W_{\tilde{\phi}\tilde{\phi}}(\tilde{\phi}^{(0)})\,=\,\frac{V_{\phi\phi}\biggl(e^{\frac{\tilde{\phi}^{(0)}}{\sqrt{2\omega_0+3}}}\biggr)}{2\omega_0+3}\,=\,\frac{V_{\phi\phi}(\phi^{(0)})}{2\omega_0+3}$ and by the definition of mass ${m_\phi}^2$, given in Eq. (\ref{mass_defin}),  we obtain $W_{\tilde{\phi}\tilde{\phi}}(\tilde{\phi}^{(0)})\,=\,{m_\phi}^2/\phi^{(0)}$. Finally,  the energy-momentum tensor $\tilde{T}_{\mu\nu}$ is given by the following expression

\begin{eqnarray}
\tilde{T}_{\mu\nu}\,=\,\rho\exp\left(-\frac{2\,\tilde{\phi}}{\sqrt{2\,\omega_0+3}}\right)\tilde{u}_\mu\tilde{u}_\nu
\end{eqnarray}
where $\tilde{g}_{\mu\nu}\tilde{u}^\mu\tilde{u}^\nu\,=\,1$ then $\tilde{u}_0\,=\,\sqrt{\phi^{(0)}+2\,\tilde{\Phi}}$. In the Newtonian limit, we find $\tilde{T}^{(1)}_{00}\,=\,\rho/\phi^{(0)}$ and $\tilde{T}^{(1)}\,=\,\rho/{\phi^{(0)}}^2$. It remains only to calculate the source term $\delta\tilde{\mathcal{L}}_m/\delta\tilde{\phi}$  of the scalar field $\tilde{\phi}^{(1)}$. From the third line of (\ref{transconfTS}) and, by using the transformation rules (\ref{trans_rule}), we find the  coupling between the scalar field and the ordinary matter 

\begin{eqnarray}\label{var_matter}
\frac{\delta\tilde{\mathcal{L}}_m}{\delta\tilde{\phi}}\,&=&\,\frac{\delta}{\delta\tilde{\phi}}\biggl\{e^{-\frac{2\tilde{\phi}}{\sqrt{2\omega_0+3}}}\mathcal{L}_m\biggl(e^{-\frac{\tilde{\phi}}{\sqrt{2\omega_0+3}}}\,\tilde{g}_{\rho\sigma}\biggr)\biggr\}\,=\,\frac{-2\,e^{-\frac{2\,\tilde{\phi}}{\sqrt{2\omega_0+3}}}}{\sqrt{2\omega_0+3}}\,\mathcal{L}_m(\cdot)+e^{-\frac{2\,\tilde{\phi}}{\sqrt{2\omega_0+3}}}\,\frac{\delta\mathcal{L}_m(\cdot)}{\delta g^{\mu\nu}}\,\frac{\delta g^{\mu\nu}}{\delta \tilde{\phi}}
\nonumber\\\nonumber\\
&=&\frac{-2\,e^{-\frac{2\,\tilde{\phi}}{\sqrt{2\omega_0+3}}}}{\sqrt{2\omega_0+3}}\,\mathcal{L}_m(\cdot)+e^{-\frac{2\,\tilde{\phi}}{\sqrt{2\omega_0+3}}}\,\frac{\mathcal{L}_m(\cdot)}{2}(g_{\mu\nu}-u_\mu u_\nu)\,\frac{\tilde{g}^{\mu\nu}\delta\,\phi(\tilde{\phi})}{\delta\,\tilde{\phi}}
\\\nonumber\\
&=&\nonumber\frac{-2\,e^{-\frac{2\,\tilde{\phi}}{\sqrt{2\omega_0+3}}}}{\sqrt{2\omega_0+3}}\,\mathcal{L}_m(\cdot)+e^{-\frac{2\,\tilde{\phi}}{\sqrt{2\omega_0+3}}}\,\frac{3\,\mathcal{L}_m(\cdot)}{2}\frac{1}{\sqrt{2\omega_0+3}}\,=\,-\frac{1}{2}\frac{e^{-\frac{2\,\tilde{\phi}}{\sqrt{2\omega_0+3}}}}{\sqrt{2\omega_0+3}}\,\mathcal{L}_m(\cdot)\,=\,-\frac{e^{-\frac{2\,\tilde{\phi}}{\sqrt{2\omega_0+3}}}}{\sqrt{2\omega_0+3}}\,\rho
\end{eqnarray}
Then the system of Eqs. (\ref{NL_fieldequation_ST_EF}) becomes

\begin{eqnarray}\label{NL_fieldequation_ST_EF_2}
&&\triangle\tilde{\Phi}\,=\,\frac{\mathcal{X}\,\rho}{2}\,\,\,\,\,\,\,\,\,\,\,\,\,\,\,\,\tilde{\Psi}\,=\,\tilde{\Phi}\nonumber
\\\\
&&\biggl[\triangle-{m_\phi}^2\biggr]\tilde{\phi}^{(1)}\,=\,-\frac{\mathcal{X}\,\rho}{\phi^{(0)}\sqrt{2\omega_0+3}}
\nonumber
\end{eqnarray}
and their solutions in the case of pointlike source are

\begin{eqnarray}\label{NL_fieldequation_ST_EF_2_sol}
\tilde{\Phi}\,=\,-\frac{ G\,M}{|\textbf{x}|}\,\,\,\,\,\,\,\,\,\,\,\,\,\,\,\,\tilde{\Psi}\,=\,\tilde{\Phi}\,\,\,\,\,\,\,\,\,\,\,\,\,\,\,\,\tilde{\phi}^{(1)}\,=\,\frac{1}{\phi^{(0)}\sqrt{2\omega_0+3}}\frac{r_g}{|\textbf{x}|}\,e^{-m_\phi|\textbf{x}|}
\end{eqnarray}
The difference in Eqs.(\ref{diff_pot}) between the gravitational potentials is satisfied by using the expression of scalar field in the Jordan frame (first line of (\ref{NL-solution_ST})) where, obviously, we set $\omega(\phi)\,=\,-\omega_0/\phi$. In fact we find

\begin{eqnarray}
\tilde{\Phi}-\phi^{(0)}\Phi\,=\,\frac{GM}{2\,\omega_0+3}\frac{e^{-m_\phi |\textbf{x}|}}{|\textbf{x}|}\,=\,\frac{\phi^{(1)}}{2}
\end{eqnarray}
and an analogous relations is found also for the couples $\Psi,\,\tilde{\Psi}$. Furthermore we can check  also the transformation rules (\ref{trans_rule}) and (\ref{trans_rule_pert}) for the solutions (\ref{NL-solution_ST}) and (\ref{NL_fieldequation_ST_EF_2_sol}) of the scalar fields $\phi,\,\tilde{\phi}$.

The redefinition of the gravitational constant $G$ (as performed in the Jordan frame $G\,\rightarrow\,G^*$ in the case of Brans-Dicke theory \cite{brans}) is not available when we are interested to compare the outcomes in both frame. In fact the couple of potentials $\Phi, \tilde{\Phi}$ differs not only from the dynamical contribution of the scalar field ($\phi^{(1)}$) but also from the definition of the gravitational constant. Furthermore,  in the Einstein frame, the scalar field $\tilde{\phi}$ does not contribute (the coupling constant between $R$ and $\tilde{\phi}$ is vanishing), then we find the same outcomes of General Relativity with ordinary matter. However by supposing the Jordan frame as starting point and  coming back via conformal transformation, we find that the gravitational constant is not invariant and depends on the background value of the scalar field in the Einstein frame, that is  $G\,\rightarrow\,G_{eff}\,\propto\,e^{-\tilde{\phi}^{(0)}}G$.

\section{The case of $f(R)$-gravity}

Recently,  several authors claimed that higher-order theories of gravity and among them,  $f(R)$ gravity, are characterized by an ill defined behavior in the Newtonian regime. In particular,   it is discussed  that Newtonian corrections of the gravitational potential violate experimental constraints since these quantities can be recovered by a direct analogy with Brans-Dicke gravity  simply supposing the Brans-Dicke characteristic parameter $\omega_0$ vanishing (see \cite{olmo} for a discussion). Actually, the calculations of the Newtonian limit of $f(R)$-gravity, directly performed in a rigorous manner, have showed that this is not the case \cite{cqg, stabile, stabile2, noi-prd, dick} and it is possible  to discuss also the analogy with Brans-Dicke gravity. The issue is easily overcome once the correct analogy between $f(R)$-gravity and the corresponding scalar-tensor framework is taken into account. It is worth noticing  that several results already achieved in the Newtonian regime, see e.g.\cite{hans,stelle}, are confirmed by the present   approach.

In literature, it is shown that $f(R)$ gravity models can be rewritten in term of a scalar-field Lagrangian non-minimally coupled with gravity but without  kinetic term implying that the Brans-Dicke parameter is $\omega(\phi)\,=\,0$. This fact is considered the reason for the ill-definition of the weak field limit that should be $\omega\rightarrow \infty$ inside the Solar System.

Let us  deal with the $f(R)$ gravity formalism in order to set correctly the problem. The action is  

\begin{eqnarray}
\mathcal{A}^{JF}_{f(R)}=\int
d^4x\sqrt{-g}\biggl[f(R)+\mathcal{X}\mathcal{L}_m\biggr]
\end{eqnarray}
and the field equations are

\begin{eqnarray}\label{fe}
f_RR_{\mu\nu}-\frac{f}{2}\,g_{\mu\nu}-f_{R;\mu\nu}+g_{\mu\nu}\Box
f_R\,=\,\mathcal{X}\,T_{\mu\nu}
\end{eqnarray}
with the trace
\begin{eqnarray}
3\,\Box f'+f_RR-2f\,=\,\mathcal{X}T
\end{eqnarray}
where $f_R\,=\,\frac{df}{dR}$. These equations can be recast in the framework of scalar-tensor gravity as son as  we select a particular expression for the  free parameters of  the theory. The result is the so-called O'Hanlon theory \cite{ohanlon} which can be written as
\begin{eqnarray}\label{ohanlon}
\mathcal{A}^{JF}_{OH}=\int d^4x\sqrt{-g}\biggl[\phi
R+V(\phi)+\mathcal{X}\mathcal{L}_m\biggr]
\end{eqnarray}
The field equations are obtained by starting from Eqs. (\ref{fieldequation_ST})

\begin{eqnarray}
\label{fieldequation_OH}
&&\phi\,R_{\mu\nu}-\frac{\phi\,R+V(\phi)}{2}\,g_{\mu\nu}-\phi_{;\mu\nu}+g_{\mu\nu}\Box\,
\phi\,=\,\mathcal{X}\,T_{\mu\nu}\nonumber\\\nonumber\\
&&R+V_{\phi}(\phi)\,=\,0\\\nonumber\\\nonumber
&&\phi\,R+2V(\phi)-3\,\Box\,\phi\,=\,-\mathcal{X}\,T
\end{eqnarray}
By supposing that the Jacobian of the transformation $\phi\,=\,f_R$ is non-vanishing, the two representations  can be mapped one into the other considering the following equivalence

\begin{eqnarray}
\label{equiv}
&&\omega(\phi)\,=\,0\nonumber\\\nonumber\\
&&V(\phi)\,=\,f-f_RR\\\nonumber\\\nonumber
&&\phi\,V_\phi(\phi)-2V(\phi)\,=\,f_RR-2f
\end{eqnarray}
From the definition of the mass (\ref{mass_defin}) we have $\phi\,V_\phi(\phi)-2V(\phi)\,=\,3\,{m_\phi}^2\phi^{(1)}$, then we have also $f_RR-2f\,=\,3\,{m_\phi}^2\phi^{(1)}$ and by performing the Newtonian limit on the function $f$ \cite{stabile}, we get $f_R(0)R^{(1)}\,=\,-3\,{m_\phi}^2\phi^{(1)}$. The spatial evolution of Ricci scalar is obtained by solving the field Eq.(\ref{fe})
\begin{eqnarray}\label{Ricci_sol}
R^{(1)}\,=\,-\frac{3\,{m_\phi}^2\phi^{(1)}}{f_R(0)}\,=\,-\frac{{m_\phi}^2\,r_g}{f_R(0)}\frac{e^{-m_\phi|\textbf{x}|}}{|\textbf{x}|}
\end{eqnarray}
   without using the conformal transformation \cite{stabile, stabile2}. The solution for the potentials $\Phi, \Psi$ are obtained simply by setting $\omega(\phi)\,=\,0$ in Eqs. (\ref{NL-solution_ST}) and $\phi^{(0)}\,=\,f_R(0)$. In the case $f(R) \rightarrow\,R$, from the second line of (\ref{equiv}), $V(\phi)\,=\,0\,\rightarrow\,m_\phi\,=\,0$ and the solutions (\ref{NL-solution_ST}) become the standard  s Schwarzschild solution in the Newtonian limit.

Finally,  we can consider a Taylor expansion \footnote{The terms resulting from $R^n$ with $n \geq 3$ do {\it not} contribute at the Newtonian order.} of the form $f\,=\,f_R(0)\,R^{(1)} +\frac{f_{RR}(0)}{2}\, {R^{(1)}}^2$ so that the associated scalar field reads $\phi\,=\,f_R(0)+f_{RR}(0)\,R^{(1)}$. The relation between $\phi$ and $R^{(1)}$ is $R^{(1)}\,=\,\frac{\phi-f_R(0)}{f_{RR}(0)}$ while the self-interaction potential (second line of (\ref{equiv})) turns out the be $V(\phi)\,=\,-\frac{(\phi-f_R(0))^2}{2\,f_{RR}(0)}$ satisfying the conditions $V(f_R(0))\,=\,0$ and $V_\phi(f_R(0))\,=\,0$. In relation to the definition of the scalar field, we can opportunely identify $f_R(0)$ with a constant value $\phi^{(0)}\,=\,f_R(0)$ which  justifies the previous ansatz for  matching  solutions in the limit of General Relativity. Furthermore, the mass of the scalar field can be expressed in term of the Lagrangian parameters as ${m_\phi}^2\,=\,\frac{1}{3}\phi^{(0)}\,V_{\phi\phi}(\phi^{(0)})\,=\,-\frac{f_R(0)}{3f_{RR}(0)}$. Also in this case the value of mass is the same obtained by solving the problem without invoking the scalar tensor analogy \cite{stabile,stabile2}. However with this last remark, it is  clear the analogy between $f(R)$-gravity and a particular class of scalar tensor theories \cite{ohanlon}. 

\section{Discussion and Conclusions}
The debate of selecting a physical frame by conformal transformations has become pressing in relation to the problem of cosmological dark components. In fact, both material and geometrical origin of such dark effects are today valid and discrimination could be, in some sense, related to the selection of a set of physical quantities that are conformally invariant, a part the discovery of some new ingredient at fundamental level.

Besides this issue, there is the general problem to understand how the gauge group and the conformal group are related in a given theory of gravity. Actually, the gauge invariance breaks in the weak field limit but some conformal quantities could be preserved contributing in the selection of the physical frame.

In  this paper, we have taken into account the problem of weak field limit of scalar-tensor theories of gravity showing how the Newtonian limit behaves in the Jordan and in the Einstein frame. The general result is that  Newtonian potentials, masses and other physical quantities can be  compared in both frames once the perturbative analysis is performed.  The main point is that  if such an analysis is carefully developed in the same frame,  the perturbative process can be controlled  step by step leading to coherent results in both frames. In other words, also if the gauge invariance is broken, there is the possibility to control conformal quantities.  In particular, it is important to fix the relation between conformally related potentials in order to understand how gravitational coupling and Yukawa-like corrections behave. Specifically, the  potentials 
\begin{eqnarray}
&&\Phi(\textbf{x})\,=\,-\frac{GM}{\phi^{(0)}|\textbf{x}|}\biggl\{1-\frac{e^{-m_\phi |\textbf{x}|}}{2\,\omega(\phi^{(0)})\,\phi^{(0)}-3}\biggr\}\qquad
\Psi(\textbf{x})\,=\,-\frac{GM}{\phi^{(0)}|\textbf{x}|}\biggl\{1+\frac{e^{-m_\phi |\textbf{x}|}}{2\,\omega(\phi^{(0)})\,\phi^{(0)}-3}\biggr\}\nonumber\\\\&&\qquad\qquad\qquad\qquad\phi(\textbf{x})\,=\,\phi^{(0)}-\frac{1}{2\,\omega(\phi^{(0)})\,\phi^{(0)}-3}\frac{r_g}{|\textbf{x}|}\,e^{-m_\phi |\textbf{x}|}\nonumber
\end{eqnarray}
achieved in the Jordan frame (see Eqs.(\ref{NL-solution_ST})  can be rigorously compared with their counterparts in the Einstein frame 
\begin{eqnarray}
\tilde{\Phi}\,=\,-\frac{ G\,M}{|\textbf{x}|}\,\,\,\,\,\,\,\,\,\,\,\,\,\,\,\,\tilde{\Psi}\,=\,\tilde{\Phi}\,\,\,\,\,\,\,\,\,\,\,\,\,\,\,\,\tilde{\phi}\,=\,\sqrt{2\omega_0+3}\,\ln\phi^{(0)}+\frac{1}{\phi^{(0)}\sqrt{2\omega_0+3}}\frac{r_g}{|\textbf{x}|}\,e^{-m_\phi|\textbf{x}|}
\end{eqnarray}
see Eqs.(\ref{NL_fieldequation_ST_EF_2_sol}) when we set $\tilde{\omega}(\tilde{\phi})\,=\,-1/2$, $\Xi\,=\,1$ and $\omega(\phi)\,=\,-\omega_0/\phi$. This is the main result of this paper which, in principle, could constitute a paradigm to compare physical quantities in both frames.
In this sense, the observable consequences of conformal transformations can be achieved.
In a forthcoming paper, we will discuss how to give experimental constraints to these results.

\section*{Acknowledgements}

The authors  would like to thank the anonimous Referee  for
useful  comments and suggestions that allowed to improve the paper.

\end{document}